\documentclass[a4paper,pre,twocolumn,longbibliography]{revtex4-1}
\usepackage{graphicx}
\usepackage{amsmath}
\usepackage{amssymb}
\usepackage{epstopdf}
\usepackage{natbib}
\usepackage[normalem]{ulem}
\usepackage{xcolor}
\usepackage{cancel}

\newcommand \beq{\begin{equation}}
\newcommand \eeq{\end{equation}}
\newcommand{\upd}{\mathrm{d}}

\newcommand{\lbar}{\bar{\lambda}}
\newcommand{\tauc}{t_c}
\newcommand{\xast}{x_\ast}
\newcommand{\hTi}{h_{\mathrm{Ti}}}
\newcommand{\ETi}{E_{\mathrm{Ti}}}
\renewcommand{\Gamma}{\Pi}

\definecolor{MSLightBlue}{rgb}{.31,.506,.741}
\definecolor{MSDarkBlue}{rgb}{.51,.31,.741}

\begin{document}

\title{Lubricated wrinkles: imposed constraints affect the dynamics of wrinkle coarsening}

\author{Ousmane Kodio, Ian M.~Griffiths and Dominic Vella}
\email[]{dominic.vella@maths.ox.ac.uk}
\affiliation{ Mathematical Institute, University of Oxford, Woodstock Rd, Oxford, OX2 6GG, UK}

\begin{abstract}
We study the dynamic coarsening of wrinkles in an elastic sheet that is compressed while lying on a thin layer of viscous liquid. When the ends of the sheet are instantaneously brought together by a small distance, viscous resistance initially prevents the sheet from adopting a globally buckled shape. Instead, the sheet  accommodates the compression by wrinkling. Previous scaling arguments suggested that a balance between the sheet's bending stiffness and viscous effects lead to a wrinkle wavelength $\lambda$ that increases with time $t$ according to $\lambda\propto t^{1/6}$. We show that taking proper account of the compression constraint leads to a logarithmic correction of this result, $\lambda\propto (t/\log t)^{1/6}$. This correction is significant over experimentally observable time spans, and leads us to reassess previously published experimental data.
\end{abstract}

\maketitle

\section{Introduction}
\label{sec:introduction}

The term `fluid--structure interaction' is usually used to describe the interaction between large scale structures (such as bridges and aircraft) and high-speed flow \cite{Paidoussis2011}. Motivated by applications at small scales, including Microelectromechanical systems (MEMS) and the locomotion of microscopic organisms \cite{Lauga2016}, however, there has recently been increased interest in fluid--structure interaction at low Reynolds number \cite{Duprat2016}. Of this general class of problems, those involving the coupling between the elastic deformation of a slender structure, such as a beam, and flow in a thin, viscous layer allow the essential interaction between elasticity and hydrodynamics to be teased out relatively easily:  the pressure jump across the beam can be related to its shape (via the Euler--Bernoulli beam equation) and used to develop model equations for the deflection of the beam's centre-line. The resulting models are then amenable to analytical, as well as numerical, techniques. Furthermore, the results of these models are useful in applications at a range of scales from microfluidic devices that incorporate elastic elements \cite{Hosoi2004,Holmes2013}, through soft robots \cite{Matia2015} to deformations on a geological scale \cite{Budd2000,Michaut2011,Lister2013}. In these studies, the focus is, quite naturally, the effect of the beam's bending stiffness on the resulting dynamics. However, elastic beams can also support a tensile or compressive force along the axis. While in many situations of interest, one end of the beam is free \cite{Hosoi2004,Tulchinsky2016a}, making the neglect of this force entirely appropriate, in other situations the elastic beam is subject to some confinement: both bending and compression may play a role.

The interaction between a compressive force and a thin object's bending stiffness is known to be an  intricate one. In the
simplest possible case, compressing an elastic beam, the beam buckles once the compressive force reaches a critical value. This
process, known as Euler buckling \cite{Landau1960}, generally leads to the development of a single bump, occupying the whole
system. With additional physics, however, an intermediate length scale may be selected, leading to the development of an array
of regular wrinkles. For example, an elastic sheet floating on a deep liquid bath and compressed quasi-statically wrinkles with
a wavelength that is determined by the balance between the sheet's bending stiffness and the hydrostatic pressure within the
liquid \cite{Pocivavsek2008}. The dynamic buckling of an elastic beam immersed in a liquid and subject to a constant compressive
force $P$ has been studied by Biot \cite{biot1957folding}; here wrinkles form with a wavelength that is proportional to ~$P^{-1/2}$.

As well as being visually striking, the regular patterns formed by wrinkles have many potential applications including photonic devices \cite{Kim2012,Bayley2014} and surfaces with anisotropic wetting properties \cite{Chung2007}, amongst others. In general, these wrinkle patterns are  determined statically by properties such as the thickness of the beam and the contrast in elastic stiffnesses between the substrate and the beam. This quasi-static picture limits the range of applications somewhat, for example preventing the development of `chirp' in photonic devices. What is required in such scenario  is a wavelength that evolves in time in a controllable way \cite{Leocmach2015hierarchical}. In other scenarios, wrinkles are an intermediate step caused in the manufacture of devices, and so it is the time scale over which they disappear that is of most interest \cite{Huang2002}.

\begin{figure}
  \centering
    \includegraphics[width=0.9\columnwidth]{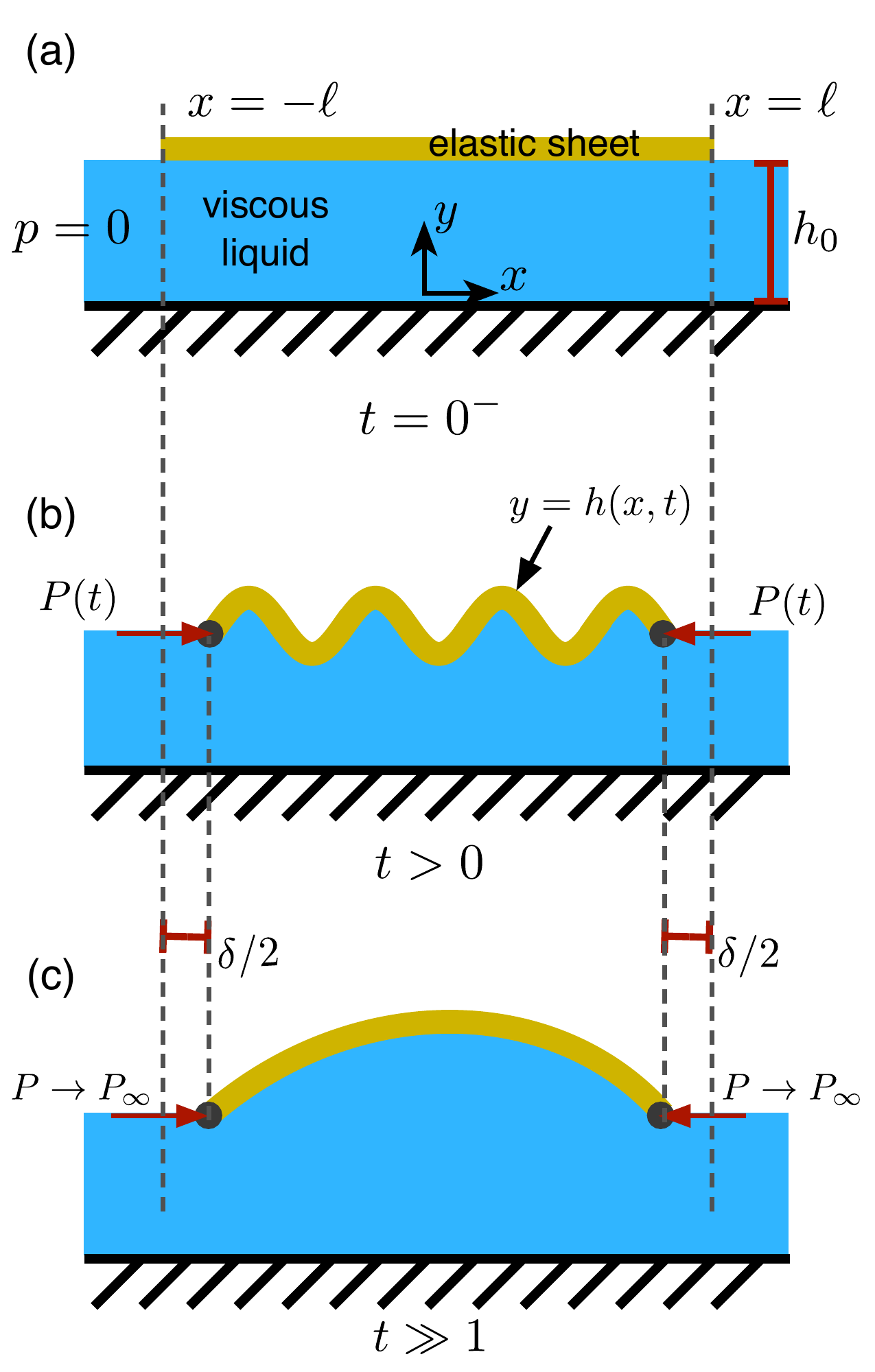}
    \caption{The dynamic wrinkling of an elastic sheet on a thin viscous film. (a) Initially an elastic sheet
      of length $2\ell$ is supported on a viscous film of thickness $h_0$. (b) At time $t=0^+$ an end-shortening $\delta/2$ is imposed (instantaneously) to each of the two ends, and maintained for all $t>0$. (The
      imposed displacement requires a compressive force $P$, per unit width, to be applied at the ends; $P$ is
      not known \emph{a priori} and may evolve dynamically.) The sheet accommodates this displacement, while preserving its contour length by buckling. Relatively large amplitude displacements require viscous fluid to be drawn in, which is a slow process. At intermediate times, the sheet adopts a wrinkled profile, with fine wrinkles that coarsen as more liquid is sucked in. (c) Ultimately, the sheet adopts the lowest Euler-buckling mode.}
  \label{fig:1}
\end{figure}

A simple form of dynamics may be obtained by using a  beam whose thickness increases as additional material is
polymerized \cite{Bayley2014}. However, several papers have focussed on experiments in which a relatively
stiff layer (generally metal) is adhered to a soft polymer layer. This composite is then  heated above the glass transition temperature of the polymer layer \cite{Yoo2003,Yoo2005,Vandeparre2010}. The differential thermal expansion is believed to be the cause of the observed wrinkles. However, since the substrate is now rubbery, it is able to flow, and the wrinkles gradually coarsen. Here, the differential thermal expansion is merely a way by which the top layer is forced to adopt a longer contour length than the lower layer, becoming relatively compressed. The essential mechanism is  illustrated in fig.~\ref{fig:1} without the complication of heating: an elastic sheet lies on a thin layer of a viscous liquid, of thickness $h_0$(fig.~\ref{fig:1}a). When its two ends are brought together by a fixed distance $\delta$ (the end-shortening), the sheet buckles to maintain its natural length. However, adopting the Euler-buckling profile (fig.~\ref{fig:1}c) would require a great deal of viscous fluid to be brought in to the system, which cannot happen instantaneously. Instead, the sheet adopts a wrinkled profile (fig.~\ref{fig:1}b), allowing the length constraint to be met with only minimal movement of viscous liquid. Over time, these wrinkles coarsen, until eventually the system does indeed adopt the expected Euler-buckling profile. Our aim in this paper is to go beyond the scaling analysis presented previously \cite{Vandeparre2010,Leocmach2015hierarchical} and to account for the global nature of the constraint appropriately.

\section{Theoretical formulation}

We assume that the typical thickness of the viscous film, $h_0$, is small compared with the horizontal extent of
the system, $\ell$. Assuming also that the horizontal length scale on which the film thickness varies is large
compared with the thickness (i.e.~the angle of deflection of the beam is small), we may use the lubrication
approximation \cite{Leal2007} to describe the thickness profile; in particular, the evolution of the film
thickness $h(x,t)$ is related to the pressure profile within the film, $p(x,t)$, by Reynolds' equation \cite{Leal2007}
\beq
\frac{\partial h}{\partial t} = \frac{1}{12 \mu} \frac{\partial }{\partial x} \left( h^{3}\frac{\partial
    p}{\partial x} \right),
  \label{eqn:Reynolds}
  \eeq 
where $\mu$ is the viscosity of the liquid inside the film.
Assuming that motion occurs quickly enough that the beam is
  instantaneously in equilibrium (i.e.~neglecting inertia) the
  pressure in the film is given by the beam equation \cite{Landau1960}
\begin{equation}
  p = B \frac{\partial^{4} h }{\partial x^{4}} +P\frac{\partial^{2} h
  }{\partial x^{2}},
 \label{eqn:Beam}
\end{equation} 
where $B=E h_b^3/[12(1-\nu^2)]$, is the bending stiffness of the beam (with $E$ its Young's modulus,
$h_b$ its thickness, and $\nu$ its Poisson ratio), while $P$ is the compressive force applied at the ends
of the beam. Note that here our use of the linear beam equation is consistent with the lubrication
approximation already made. Furthermore, the compressive force $P$ is homogeneous, i.e.~$P=P(t)$: variations
in $P$ with $x$ arise from viscous stresses but are negligible. To see this, we use a horizontal force balance, which gives $\partial P/\partial x=\mu \partial u/\partial y\vert_{y=h}$ where
$u$ is the horizontal velocity, with $y$  the vertical coordinate. Since the flow is Poiseuille, we have
$\partial P/\partial x \approx (\upd p/\upd x) h/2 \sim h_0 \Delta p/\ell$, where $\Delta p$ is the typical
hydrodynamic pressure difference over the length of the beam. Hence the typical change in the compressive
force within the sheet due to fluid shear stresses is $\Delta P\sim h_0\Delta p$. The typical pressure change
in the viscous film $\Delta p\sim Bh_0/\ell_\ast^4$, so that $\Delta P\sim B h_0^2/\ell_\ast^4$ (here
$\ell_\ast$ is a relevant horizontal length scale, which may change during the evolution but will always
satisfy $\ell_\ast\lesssim \ell$).  Finally, we note that the balance between the first and second terms on the
RHS of \eqref{eqn:Beam} suggests that $P\sim B/\ell_\ast^2$, and so we conclude that
$\Delta P \sim B h_0^2/\ell_\ast^4\ll B/\ell_\ast^2\sim P$ by virtue of the thin-layer approximation
$h_0/\ell \ll 1$. As a result we conclude that spatial variations in the compressive force $P$ may be
neglected, so that $P(x,t)\approx P(t)$. We note that in neglecting spatial variations in $P$, our approach differs from previous numerical work \cite{Huang2002,Im2005,Huang2006}.

We therefore find that the film thickness is governed by
\begin{align}
  \frac{\partial h}{\partial t} = \frac{B}{12\mu}\frac{\partial }{\partial x} \left[h^{3}\left(\frac{\partial^5
  h}{\partial x^5}  + \frac{P}{B} \frac{\partial^3
  h}{\partial x^3}\right) \right] 
\label{eq:pdedim}
\end{align} for $t>0$, $-\ell<x<\ell$.

The motion studied in this paper is driven by an imposed end--end compression $\delta$.  Assuming that the beam is inextensible (which corresponds to an assumption of being sufficiently slender \cite{Landau1960,Pandey2014}), this end--end compression imposes an integral constraint on the problem. With the approximation of small slopes, this constraint may be written \begin{equation}
  \label{eqn:Constraint}
 \tfrac{1}{2} \int_{-\ell}^{\ell} \left(
\frac{\partial h}{\partial x} \right)^{2}~\upd x   = \delta.
\end{equation}

\subsection{Initial and boundary conditions}

The problem \eqref{eq:pdedim} subject to the constraint \eqref{eqn:Constraint} requires an initial condition for $h$ and six boundary conditions. We denote the initial shape of the beam by \begin{equation}
  h(x,t=0) = f(x)
 \end{equation} for some given function $f(x)$.
  We also assume that the film thickness is prescribed at the edges, giving the boundary conditions
  \begin{equation} 
  h(x=-\ell,t) = h(x=\ell,t)= h_{0}.
\end{equation} 
Various further boundary conditions are possible for beams (e.g.~no shear force, no torque, or clamped). For simplicity, we
shall assume that no moment is applied to the beam at its ends (so that $h_{xx}$ vanishes there) and that the pressure is
atmospheric (without loss of generality zero) there too; because of the beam equation \eqref{eqn:Beam}, this guarantees that $h_{xxxx}=0$ at these edges. We therefore have
  \begin{align}    
 & h_{xx}(x=-\ell,t) = h_{xx}(x=\ell,t)= 0, \\
 &h_{xxxx}(x=-\ell,t) = h_{xxxx}(x=\ell,t)= 0.
  \end{align}


\subsection{Scaling analysis\label{sec:scaling}}

To gain some understanding of the evolution of the film thickness $h(x,t)$, as described by \eqref{eqn:Reynolds} and
\eqref{eqn:Beam}, we begin by neglecting the compressive force, i.e.~we set $P(t)\equiv0$. For small variations of the film
thickness from the uniform value $h_0$, we see that, in scaling terms, we have \beq \frac{\partial h}{\partial t}\approx
\frac{Bh_0^3}{12\mu}\frac{\partial^6h}{\partial x^6}.  \eeq This linear equation has similarity
solutions \cite{FLITTON2004,Tulchinsky2016a,Arutkin2016} in which the horizontal length scale
$x \propto (Bh_0^3/\mu)^{1/6}t^{1/6}$. We therefore anticipate that the observed wrinkle wavelength should coarsen with time
according to \beq \lambda(t)\sim \left(\frac{B h_0^3}{\mu}t \right)^{1/6}.
\label{eqn:NaiveScaling}
\eeq

The scaling law \eqref{eqn:NaiveScaling} is identical to that given by energy considerations
\cite{Vandeparre2010}. With a non-zero compressive force, the similarity structure of the problem appears to remain if
$P(t)\propto (B^2\mu/h_0^3)^{1/3}t^{-1/3}$. We shall see shortly that neglecting the constraint \eqref{eqn:Constraint}
is, in fact, an over-simplification in the problem considered here. First, however, we consider the appropriate
non-dimensionalization of our problem.

\subsection{Non-dimensionalization}

It is natural to scale the thickness of the film with the value that it takes at the edges, $h_0$. There is no natural
horizontal length scale that characterizes the wrinkling behaviour in the problem (hence the appearance of similarity solutions in the unconfined case). We therefore
introduce an arbitrary horizontal scale, $\xast$, which leads to natural time and force scales \beq t_{*} = \frac{12 \mu \xast^{6} }{Bh_0^{3} }, \quad P_{*} = \frac{B}{\xast^{2}}.  \eeq Introducing dimensionless variables \beq \tilde{t}
=t/t_{*},\quad \tilde{x}=x/\xast , \quad \tilde{h} = h/h_{*}, \quad \tilde{P}= P/P_{*},  \eeq 
 (and immediately dropping tildes) we
find that \eqref{eq:pdedim} becomes \beq \frac{\partial h}{\partial t} = \frac{\partial }{\partial x}
\left[h^{3}\left(\frac{\partial^5 h}{\partial x^5} + P \frac{\partial^3 h}{\partial x^3}\right) \right]
\label{eq:pdend}
\eeq for $t>0$, $-L<x<L$, where $L=\ell/\xast$.

In the problem as currently specified, there are two sources of nonlinearity. The first is the integral
constraint corresponding to the imposed end-shortening, \eqref{eqn:Constraint}. This nonlinearity arises from the geometry, and
so we shall refer to it as the \emph{geometric nonlinearity}. The second source of nonlinearity is the nonlinear permeability
that arises in Reynolds' equation, i.e.~the $h^3$ terms in \eqref{eq:pdend}; we refer to this as the \emph{hydrodynamic
  nonlinearity}. In what follows it will be useful for us to be able to isolate the effect of the geometric, rather than
hydrodynamic, nonlinearity. To facilitate this, we let $h = 1+ u(x,t)$ so that the leading-order equation for $u$ when $|u|\ll 1$ is a linear PDE, subject to a nonlinear constraint. With this substitution the fully nonlinear problem
\begin{align}
  &\frac{\partial u}{\partial t} = \frac{\partial }{\partial x} \left(h^{3}\frac{\partial^5
    u}{\partial x^5} \right) +P(t) \frac{\partial }{\partial x} \left(h^{3}\frac{\partial^3
    u}{\partial x^3} \right)\label{eq:systembegin}\\
  &\frac{1}{2} \int_{-L}^L \left( \frac{\partial u}{\partial x} \right)^2 \upd x =\Delta, \label{eq:constraintND}\\
  &u(x,t=0) = u_{0}(x),\\
  &u(x=-L,t) = u(x=L,t) =0,\\
  &u_{xx}(x=-L,t) =u_{xx}(x=L,t) =0,\\
  &u_{xxxx}(x=-L,t) =u_{xxxx}(x=L,t) =0.
\label{eq:systemend}
\end{align}
Here the dimensionless end-shortening $\Delta = \delta x_{*}/h_{0}^{2}$.

\section{Numerical Results}
\label{sec:results}

We solve the problem specified in equations \eqref{eq:systembegin}--\eqref{eq:systemend} numerically using the method of
lines \cite{Wouwer2014}. The partial differential equation \eqref{eq:systembegin} is discretized using finite
differences in space (written in flux conservative form), leading to a series of ordinary differential equations for the
evolution of $u_i(t)=u(x_i,t)$. The constraint \eqref{eq:constraintND} is an algebraic condition on the evolution, which
we differentiate with respect to time to reduce the index of the system \cite{Peletier2001}.  The resulting
differential algebraic equation can then be integrated forward in time using the MATLAB\textsuperscript{\textregistered}
ODE solvers \cite{Wouwer2014}. Further details of the numerical scheme are given in
  Appendix \ref{app:numerics}.

In the simulations reported here we use a dimensionless system   $L=1000$ and  $N=1024$ uniformly  distributed grid points.  The numerical scheme implemented in this way runs quickly on a laptop computer (simulations reported here typically  complete in a few minutes).

Throughout the simulations, we use two types of initial conditions. The first type is a fully random noise that is uniformly  distributed.
This initial condition is then of the form \beq u(x_i,0)=u_0(x_i)= {\cal R}_i\label{eqn:RandInit} \eeq where the ${\cal R}_i$ are randomly drawn from the uniform distribution $[-\varepsilon,\varepsilon]$. 
 The second type of initial condition is localized, and of the form $u(x_{i},0) =\varepsilon \mathrm{e}^{-x_{i}^2}$. 
In both cases, we typically take $\varepsilon=10^{-2}$ here. The value of the end-shortening constraint at later times is imposed to be that of the initial condition. 

  The key quantity of interest in this study is the evolution of the average wrinkle wavelength, $\lbar(t)$. Here we
  measure $\lbar$ in a way that mimics experimental procedures \cite{Huang2007}: the number of peaks in the instantaneous beam profile, $n(t)$, is counted so that, by definition, $\lbar(t)=2L/n(t)$. We prefer this technique to others (such as discrete Fourier Transforms) since we shall see that there is not a true wavelength (in the sense of a length scale over which the pattern repeats).

\begin{figure}
  \centering
 \includegraphics[width=0.9\linewidth]{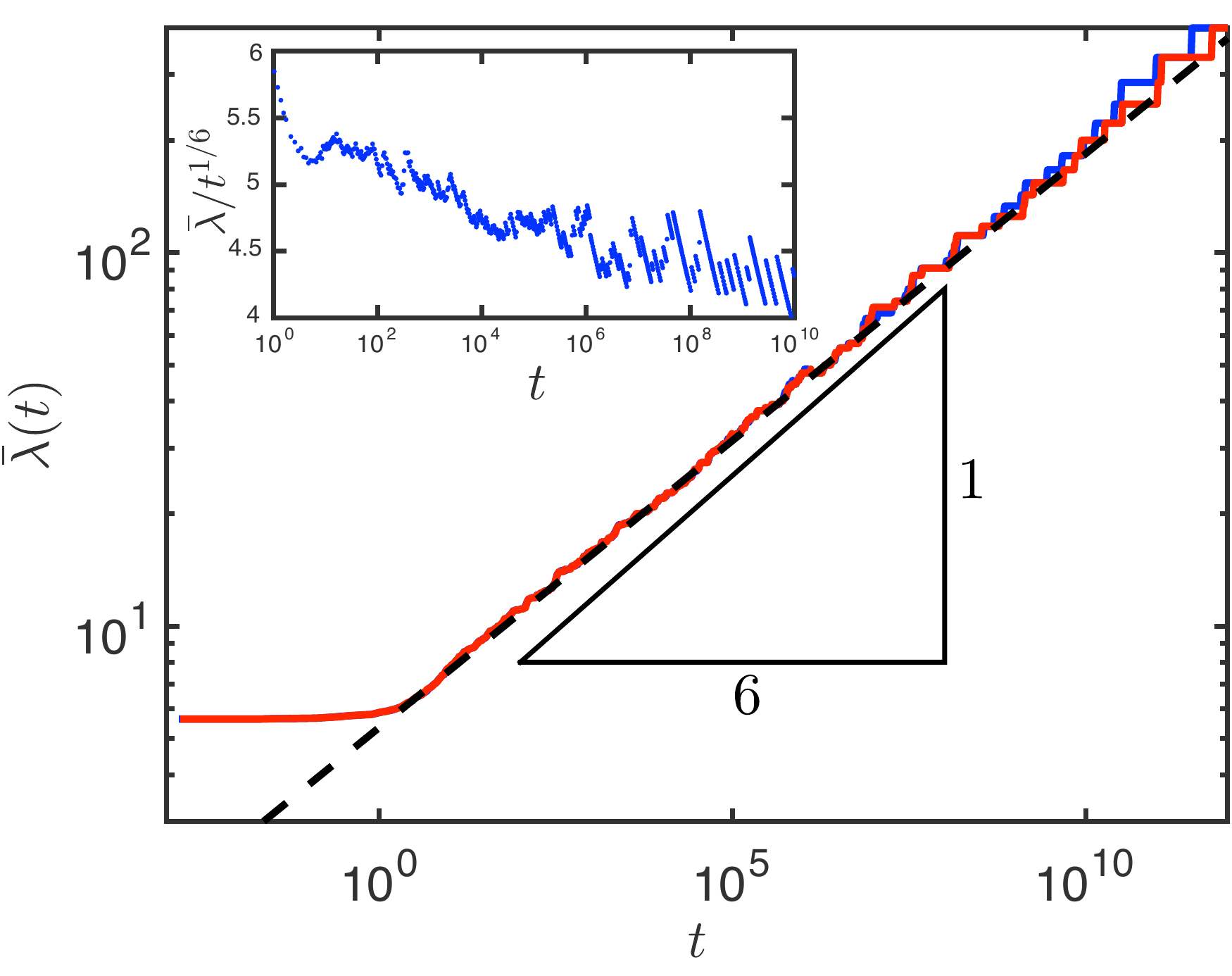}
\caption{Evolution of the mean wrinkle wavelength, $\lbar(t)$, as a function of time, determined from the
  numerical solution of \eqref{eq:systembegin}--\eqref{eq:systemend}. Main figure: The mean wavelength as determined in both the  nonlinear system (blue curve) and the linearized system (red
  curve). For times $1\lesssim t\lesssim 10^{10}$ the behaviour appears to follow a power-law relationship,
  but with $\lbar\propto t^{0.153}$ (black dashed line), which is subtly different from the power law
  $\lambda\sim t^{1/6}$ suggested previously \cite{Vandeparre2010}.  Inset: A compensated plot of the evolution of $\bar{\lambda}/t^{1/6}$ shows a systematic drift from the scaling $\lambda\sim t^{1/6}$. Here, the wavelength is determined by counting the number of   peaks in the profile at any instant, $n(t)$, so that $\bar{\lambda}(t)=2L/n(t)$. In these simulations, $L=1000$, $\Delta=17.7\times 10^{-3}$ and $N=1024$ grid points. The initial condition is $h(x,0)=1+ u_0(x)$ with $-\varepsilon\leq u(x)\leq\varepsilon$  a uniformly distributed random number at each point and $\varepsilon=10^{-2}$; the same random condition is used for both sets of simulations.
}
  \label{fig:2}
\end{figure}

The evolution of the mean wavelength  is shown in fig.~\ref{fig:2}. Despite the expectations of equation \eqref{eqn:NaiveScaling} that $\lbar\propto t^{1/6}$, over the long duration shown in our
numerical experiments we observe a significant (and systematic)
deviation from this scaling law (see the inset of fig. \ref{fig:2}). For the data range shown in this figure  a more appropriate
power law appears to be $t^{0.153}$. However, the precise value of this alternative exponent depends on the interval of time
that is considered, suggesting that the true behaviour may not quite be a power law at all.

The $\lbar\sim t^{1/6}$ scaling predicted in \eqref{eqn:NaiveScaling} was based on a simple, linearized balance between
the terms in the governing nonlinear PDE. We therefore ask whether this discrepancy is a result of one of the two
nonlinearities in the system. To address this question, we consider numerically the hydrodynamically linear
(geometrically nonlinear) PDE that is obtained by setting $h\approx1$ in \eqref{eq:systembegin}. The evolution of
$\lbar(t)$ in this linear case is identical to the hydrodynamically nonlinear problem, at least for early and
intermediate times.

\begin{figure}
  \centering
\includegraphics[width=0.9\linewidth]{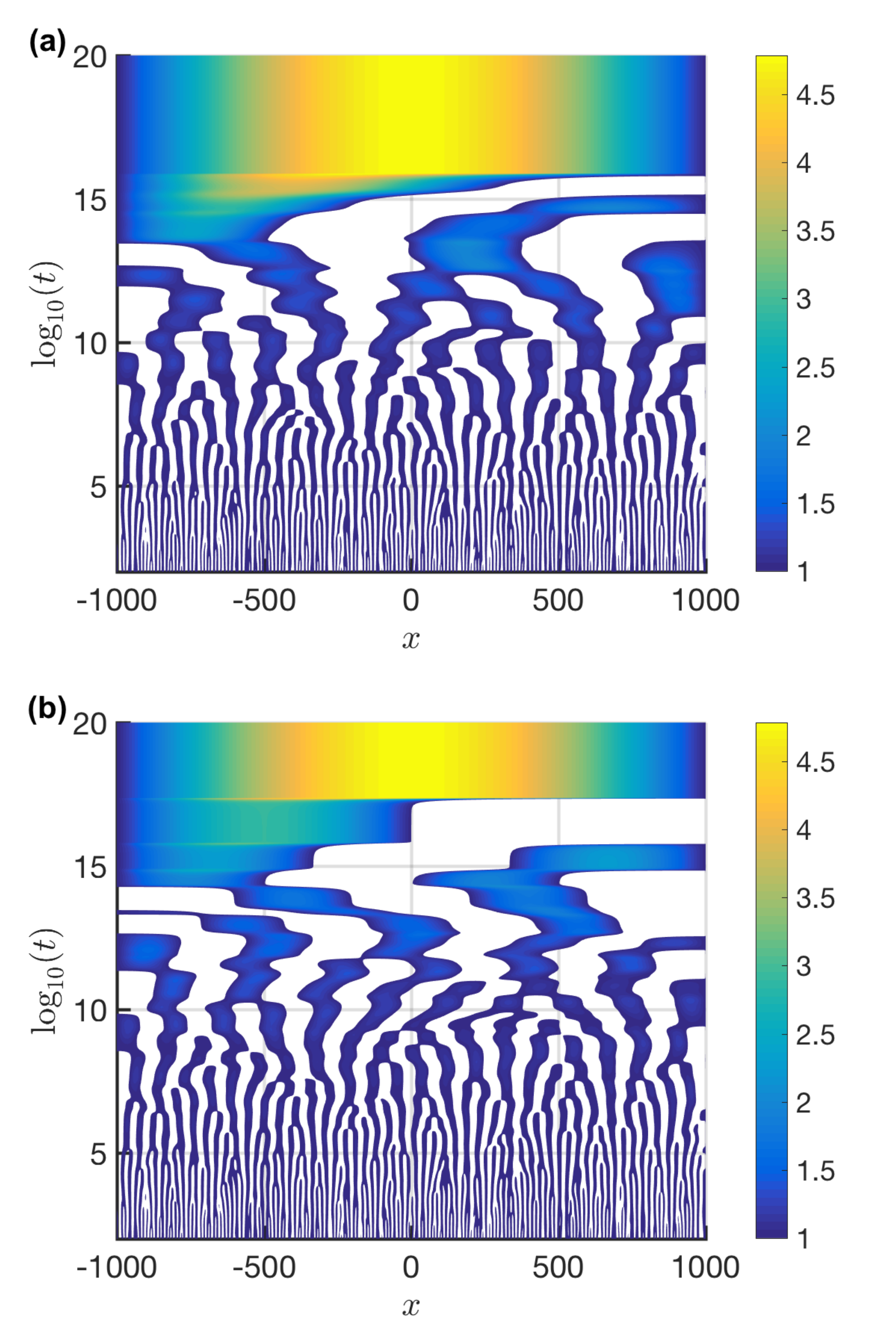}
\caption{Comparison of the evolution of the beam deflection (and hence film thickness) governed by \eqref{eq:systembegin}--\eqref{eq:systemend} observed in the (a) nonlinear and (b) linear  cases. Colours represent the value of the film thickness, $h=1+u(x,t)$, at each point of $(x,t)$-space, according to the colour bar on the right of each plot. (Note that to emphasize wrinkles, regions with $h<1$ are whited-out.) In each case, we observe qualitatively similar coarsening, motivating a more detailed analytical study of the linear problem. In these simulations,  $L=1000$, with $N=1024$ grid points and $\Delta = 17.7 \times 10^{-3}$. The initial condition  $u(x,0)$ is drawn randomly from a uniform distribution on $[-\varepsilon,\varepsilon]$ with $\varepsilon=10^{-2}$; the same random initial condition is used in both (a) and (b).
}
\label{fig:profile}
\end{figure}

The close correspondence between the linearized and nonlinear problems can also be seen in the evolution of
the film thickness $h(x,t)$, for each case, shown in fig.~\ref{fig:profile}. In particular, both the nonlinear and
linearized problems show a smoothening of the initial condition with time, and a coarsening of the
wrinkles. The similarity between the nonlinear and linearized problems suggests that the anomalous behaviour
in the wrinkle wavelength is due to the geometric nonlinearity (the integral constraint), and not the
hydrodynamic nonlinearity due to Reynolds' equation, \eqref{eq:systembegin}. To gain some insight into the
role that this geometrical nonlinearity plays, we therefore focus on the linearized case, i.e.~when the term $h^{3}$ in \eqref{eq:systembegin} is approximated by unity.   We expect that this approximation will be valid when  $u=h-1 \ll 1$, though we note that  this condition may not necessarily remain valid throughout the evolution of the system (since, as the wrinkles coarsen, $u$ increases).

\section{Analysis}

Our numerical solution of the linearized problem (shown in fig.~\ref{fig:profile}) shows that this
problem also demonstrates the wrinkle coarsening behaviour that is of interest here. More quantitatively, we
have already seen that the numerically determined wrinkle wavelength varies with time in a very similar way
(at least for small and intermediate times) in both the linearized and nonlinear problems. In particular, the
linear problem also shows that the wrinkle wavelength does not obey the expected
$\bar{\lambda}\propto t^{1/6}$ behaviour. In this section we analyse the linearized problem to gain some
understanding of this discrepancy.

To simplify the problem, we consider the spatial domain to be infinite, i.e.~$L = \infty $, which we expect to
be relevant before the wrinkles feel the effect of the edges. We then have the following linear system (albeit
with a nonlinear integral constraint) for $u(x,t)$ and the compressive force $P(t)$: \beq
\label{eq:linearbegin}
\frac{\partial u}{\partial t} = \frac{\partial^6
  u}{\partial x^6} +P(t) \frac{\partial^4u}{\partial x^4},
   \eeq with initial condition
\beq
u(x,t=0) = u_{0}(x),
\eeq boundary conditions
\beq
u=\frac{\partial^2u}{\partial x^2}=\frac{\partial^4u}{\partial x^4} =0
\eeq at $x=\pm\infty$ and the constraint
\beq
\tfrac{1}{2} \int_{-\infty}^\infty \left( \frac{\partial u}{\partial x} \right)^2~\upd x =\Delta.\\
\label{eq:linearend}
\eeq

Given that the boundary conditions are now imposed at infinity, and that the governing partial differential
equation is linear, it is natural to Fourier transform equation \eqref{eq:linearbegin}.  We adopt the following definition of the Fourier Transform:
 \begin{equation}
    \hat{u}(k,t) = \frac{1}{\sqrt{2\pi}}  \int_{-\infty}^\infty u(x,t) \mathrm{e}^{-i k x}~\upd x,
  \end{equation} so that the appropriate Inversion Theorem is
  \begin{equation}
    u(x,t) = \frac{1}{\sqrt{2\pi}}  \int_{-\infty}^\infty \hat{u}(k,t) \mathrm{e}^{i k x}~\upd k.
\label{eqn:uinv}
  \end{equation}

Taking the Fourier Transform of \eqref{eq:linearbegin} yields
\begin{equation}
  \frac{\partial\hat{u}}{\partial t} = k^{4}[P(t) - k^{2}]\hat{u},
\label{eq:spinodal}
\end{equation} which may be integrated to give
\begin{equation}
   \hat{u}(k,t) = \hat{u}_{0}(k)  \exp[-k^{6}t + \Gamma(t) k^{4}],
   \label{eqn:usoln}
\end{equation} 
where $\hat{u}_{0}(k)$ is the Fourier Transform of the initial condition and  
\begin{equation}
  \Gamma(t) = \int_{0} ^{t} P(s) \, \upd s.
\end{equation}

 Inverting the Fourier transform of the
profile by substituting \eqref{eqn:usoln} into \eqref{eqn:uinv}, we find that the solution of the problem may be expressed as
\begin{align}
  \label{eqn:uxt}
  u(x,t) = \frac{1}{\sqrt{2\pi}}\int_{-\infty}^\infty \hat{u}_{0}(k)  \exp[-k^{6}t + \Gamma(t) k^{4}]\mathrm{e}^{\mathrm{i}kx}~\upd k.
\end{align} 

We note that if the compressive force $P$ were constant then we would expect from eqn \eqref{eqn:usoln} that the solution would consist of waves with wavenumber $k=k_c=(2/3)^{1/2}P^{1/2}$, which corresponds to the fastest growing mode \cite{biot1957folding}. However, in this problem $P(t)$ evolves with time and, further, $P(t)$, and hence $\Gamma(t)$, are not known \emph{a priori}. Physically, $P(t)$ is determined by the constraint \eqref{eq:linearend} which may be rewritten in terms of $\hat{u}(k,t)$ using the Parseval--Plancherel Theorem \cite{Carleman1944}
\begin{equation}
  \label{eq:fourierconstraint}
  \frac{1}{2}\int_{-\infty}^\infty k^{2} |\hat{u}|^{2} \, \upd k = \Delta,
\end{equation}   and must hold for all time. Substituting the general solution \eqref{eqn:usoln} into the constraint \eqref{eq:fourierconstraint}, we have
\begin{align}
  \tfrac{1}{2} \int_{-\infty}^\infty k^{2} |\hat{u}_0(k)|^{2} \exp\left[-2(k^6 t
  -\Gamma(t) k^4)\right] \, \upd k = \Delta.
\label{eq:constraintFourier}
\end{align}

\subsection{Late-time  behaviour}
\label{sub-sec-late-time}

In this section we seek to determine the late-time behaviour of the compressive force $P(t)$ and the profile $u(x,t)$. Defining
$$\sigma(t) = \Gamma(t)^3 / t^{2} $$ and letting $k = t^{-1/6} \sigma(t)^{1/6} z$ 
then \eqref{eq:constraintFourier} becomes 
\begin{equation}
\label{eq:constraint-I-Delta}
\sigma^{1/2}   I\left( \sigma;t \right) = 2t^{1/2}\Delta ,
\end{equation}
where 
\begin{equation}
\label{eq:integral-Ia}
I(\sigma;t) = \int_{-\infty}^\infty z^{2} \left|\hat{u}_0\bigl(z \sigma^{1/6} t^{-1/6}\bigr)\right|^{2}  \mathrm{e}^{-2\sigma(z^{6}-z^{4})}  ~\upd z.
\end{equation} 
The natural scaling suggested in \S\ref{sec:scaling} predicts that $P\propto t^{-1/3}$, $\Gamma\propto t^{2/3}$ as
$t\to\infty$. However, this implies that $\sigma$ and $I$  both tend to constants as $t\to\infty$. This would cause the LHS of
\eqref{eq:constraint-I-Delta} to be constant, which is inconsistent with the diverging RHS. We therefore conclude that $\sigma\to\infty$, and hence that $P(t)$ must decay more slowly than $t^{-1/3}$. The quantity $\sigma$ therefore gives a measure of how far the compressive force $P$ is from the $P\sim t^{-1/3}$ behaviour required for self-similarity. To quantify the actual behaviour, however, we need to understand the asymptotic behaviour of the integral $I(\sigma;t)$ when $t\gg1$. 

This integral can                         
 be evaluated using Laplace's method \cite{Ablowitz2003}  and follows the analysis of a related problem by Budd \emph{et al.}~\cite{Budd2000}. 
 
In applying Laplace's method, we find that the dominant contribution to the integral arises when the exponent is stationary,
i.e.~when $z=z_\pm=\pm\sqrt{2/3}$. We must therefore evaluate $\hat{u}_0\bigl(z \sigma^{1/6} t^{-1/6}\bigr)$ at  $z=z_\pm$. If
the quantity $\sigma/t$ were to grow without bound then $\hat{u}_0\bigl( \sigma^{1/6} t^{-1/6}z_{\pm}\bigr)\to\hat{u}_0
(\infty)$, which, according to the Riemann-Lebesgue Lemma  is zero assuming $u_{0}(x)$ is integrable. It is also not possible
that $\sigma/t $ tends to a constant other than zero, by considering the constraint \eqref{eq:constraint-I-Delta} directly.  We therefore see that dominant contribution to the integral arises from $k\approx0$ and approximate $\hat{u}_{0}(k)\approx\hat{u}_0 (0)$. (Here, we assume that $\hat{u}_0(0)\neq0$; in Appendix \ref{app:NoFluid} we generalize the following analysis to the case $\hat{u}_0(0)=0$.)

We find that
\begin{equation}
I(\sigma;t) \sim \sqrt{\frac{\pi}{3}} |\hat{u}_0(0) |^2 \sigma^{-1/2} \mathrm{e}^{\frac{8\sigma}{27}}
 \text{ as } \sigma, t\rightarrow \infty.
 \label{eqn:IAlphaAsy}
\end{equation} and hence, upon substituting \eqref{eqn:IAlphaAsy} into \eqref{eq:integral-Ia}, that
\beq
\Delta \sim \tfrac{1}{2}\sqrt{\frac{\pi}{3}} |\hat{u}_{0}(0)|^2 t^{-1/2}\exp\left(\frac{2^{3}}{3^{3}} \sigma\right).
\label{eqn:FTconstraintAsy}
\eeq Inverting \eqref{eqn:FTconstraintAsy} we find that
\begin{equation}
\label{eq:sigma}
\sigma\sim \frac{3^{3}}{2^{3}}\left[ \tfrac{1}{2}\log(t/\tauc) \right],
\end{equation} 
 where
\beq
 \tauc=\frac{\pi}{12} \frac{|\hat{u}_{0}(0) |^4}{\Delta^{2}}.
 \label{eqn:tauc}
 \eeq 
The initial condition enters the asymptotic prediction \eqref{eq:sigma} only through $\hat{u}_0(0)$
 in the time scale $\tauc$.  Since $\hat{u}_0(0) = \tfrac{1}{\sqrt{2\pi}}\int_{-\infty} ^{\infty} u_0(x) ~\mathrm{d} x$ this quantity represents the excess (or  deficit) of fluid that is introduced in the initial condition.
 
The relative error in the leading-order expression \eqref{eqn:IAlphaAsy} may be calculated using standard arguments \cite{BenderandOrsag}.  We find that this correction, $R$, is given by 
\begin{equation}
  R = \frac{9}{64\sigma} \left[  1 +  \frac{2}{3} (\sigma/t)^{1/3} \frac{\hat{u}''_{0}(0)}{\hat{u}_{0}(0)} \right],
\end{equation}
and so conclude that for our asymptotic analysis to be valid ($R\ll 1$), we must have
\begin{equation}
  \sigma \gg 9/64  ,
\label{eq:validty-conditiona}
\end{equation} and
\begin{equation}
  t^{1/3} \sigma^{2/3} \gg \frac{3}{32} \frac{\hat{u}_{0}''(0)}{\hat{u}_{0}(0)}.
\label{eq:validty-conditionb}
\end{equation} As a consequence, our late-time analysis is valid for a symmetric localized initial condition of typical width $L_{0}$, when 
 $\log(t/t_{c}) \gg 1/12$ and $t^{1/6} (\log t)^{1/3} \gg L_0/3^{1/2} 2^{7/6}$. The first condition, \eqref{eq:validty-conditiona}, shows that $\tauc$ gives the time scale over which the system `forgets' the initial condition, $u_0(x)$; we shall comment on the physical significance of the second condition, \eqref{eq:validty-conditionb}, later. 

From the expression \eqref{eq:sigma}, we may deduce that
\begin{equation}
\label{eq:Pi1}
\Gamma(t) \sim  \frac{3}{2} t^{2/3}\left[ \tfrac{1}{2}\log \left(t/\tauc\right) \right]^{1/3}.
\end{equation}
 
Note that this asymptotic behaviour of $\Gamma(t)$ is close to the $t^{2/3}$ scaling  that would have been anticipated from the analyis in \S\ref{sec:scaling}  but includes a logarithmic correction.

With $\Gamma(t)$ determined asymptotically, we can now determine the asymptotic behaviour of the compressive force, $P(t)$, by differentiating \eqref{eq:Pi1}; we find that
\begin{equation}
\label{eq:Pasym}
P(t) \sim   \left[ \frac{\log \left(t/\tauc\right)}{2t} \right]^{1/3}.
\end{equation} 
Furthermore, for $t\gg\tauc$, the leading-order result is 
\beq
P(t) \sim  2^{-1/3} \left( \frac{\log t}{t} \right)^{1/3}.
\label{eq:P-inf}
\eeq


The effect of the constraint on the dynamics of wrinkling is embodied in the relationship \eqref{eq:Pasym} for
the compressive force that must be applied to impose the given end-shortening (after all, without this force,
there is no imposed end-shortening). However, what interests us most is the apparent wavelength of the
buckling pattern; to understand this, we now turn to the profile of the beam itself.

The inverse Fourier transform of the profile, \eqref{eqn:uxt}, may be written as
\begin{equation}
u(x,t) = \frac{1}{\sqrt{2\pi}}\frac{\sigma^{1/6}}{t^{1/6}} U_{\sigma}(\xi,t),
\end{equation} 
where 
\begin{equation}
\label{eq:-F}
U_{\sigma}(\xi,t) = \int_{-\infty}^{\infty}  \hat{u}_0\left(\frac{\sigma^{1/6}}{t^{1/6}}z\right) \exp\left[-\sigma(z^{6}-z^{4})\right] \mathrm{e}^{\mathrm{i} \xi z}~\upd z,
\end{equation} 
and $\xi=x(\sigma/t)^{1/6}$, where from equation \eqref{eq:Pi1}, we may infer the late-time behaviour of the new variable $\xi$, which is given by $\xi \sim \sqrt{\tfrac{3}{2}}\, x/(2t/\log (t/t_{c}))^{1/6} \sim x \sqrt{{3} P/2}$ as $t\rightarrow\infty$.

To evaluate $U_{\sigma}(\xi,t) $ for large $\sigma$, we again use Laplace's method; as in the evaluation of the integral in \eqref{eq:integral-Ia} we find that the integral is dominated by the behaviour around $z=z_\pm=\pm \sqrt{2/3}$. 
Making the usual approximation of the integrand for $z\approx z_\pm$, we  find that
\begin{align}
U_{\sigma}(\xi,t) \sim \sqrt{\frac{3\pi}{2}} \hat{u}_0 (0) \frac{\exp\left(\frac{4}{27}\sigma-\frac{3}{32} \frac{\xi^2}{\sigma}\right)}{\sigma^{1/2}} \cos(\sqrt{\tfrac{2}{3}} \xi),
\end{align}  for $\sigma\gg1$ and hence that the long-time behaviour of the profile may be written
\beq
u(x,t) \sim A_{0} \exp\left(-\frac{3}{32} \frac{\xi^{2}}{\sigma}\right) \cos\left(\sqrt{\tfrac{2}{3}} \xi\right),
\label{eqn:uxtasym}
\eeq
where $A_{0} \sim \pm 2^{5/6} 3^{-1/4} \pi^{-1/4} \Delta^{1/2}  (\log t )^{-1/3} t^{1/12}$. 
In \eqref{eqn:uxtasym}, the positive sign for $A_{0}$ corresponds to the case of a profile that has an initial  excess of fluid, while the negative sign corresponds to the case of a profile with an initial  deficit of fluid.
Having determined the asymptotic relationship \eqref{eqn:uxtasym}, we note two interesting features of this result. Firstly, for late times the $u(x,t)\propto t^{1/12}$ (ignoring logarithmic terms), following what would be expected from the  scaling analysis discussed in \S\ref{sec:scaling}. Secondly, the initial condition $u_0(x)$ does not appear explicitly in the result \eqref{eqn:uxtasym}: the only feature of the initial condition that is `remembered' at very late times is the imposed end-shortening, $\Delta$. We can now also interpret the second condition for the validity of the asymptotic expression,  \eqref{eq:validty-conditionb}, as the time over which the  exponential envelope in \eqref{eqn:uxtasym} becomes larger than the typical width of the initial condition, as might intuitively be expected.


The late-time profile \eqref{eqn:uxtasym} also shows that the asymptotically correct shape of the deformed beam does \emph{not} adopt a
self-similar shape: having introduced a similarity-like variable, $\xi$, in \eqref{eqn:uxtasym}, we see that there remains some (albeit weak) time dependence in the exponential term that cannot be removed by rescaling. Another perspective on this absence of a similarity solution, and the crucial role of the compressive constraint, is given in Appendix \ref{app:SimSoln}.

Finally, we note that the profile in \eqref{eqn:uxtasym} takes the form of a spatial oscillation that is modulated by an exponential decay. As such, the wrinkle pattern is not perfectly periodic and does not have a true wavelength. Nevertheless, wrinkles are observed and a natural measure of this wrinkle pattern is the distance between consecutive zeros, $\Delta x$, which is  given for late times by
\beq
\Delta x\sim2^{1/6}\pi\left[\frac{t}{\log (t/\tauc)}\right]^{1/6}.
\label{eqn:DeltaX}
\eeq 


We estimate the apparent wavelength $\lbar$ as twice the distance between zeros, hence
\beq
\lbar\sim 2^{7/6}\pi\left[\frac{t}{\log (t/\tauc)}\right]^{1/6},
\label{eqn:LambdaBarOfTtauc}
\eeq which for late times, $t\gg\tauc$, reads
\beq
\lbar\sim 2^{7/6}\pi\left(\frac{t}{\log t}\right)^{1/6}.
\label{eqn:LambdaBarOfT}
\eeq

\subsection{Comparison with numerical results}

Having determined asymptotic predictions for the evolution of the compressive force and effective wavelength evolve for late times, we now turn to compare these results with numerical simulations of the fully nonlinear problem, \eqref{eq:systembegin}--\eqref{eq:systemend}.

\subsubsection{A localized initial condition}

\begin{figure}
  \centering
\includegraphics[width=0.9\linewidth]{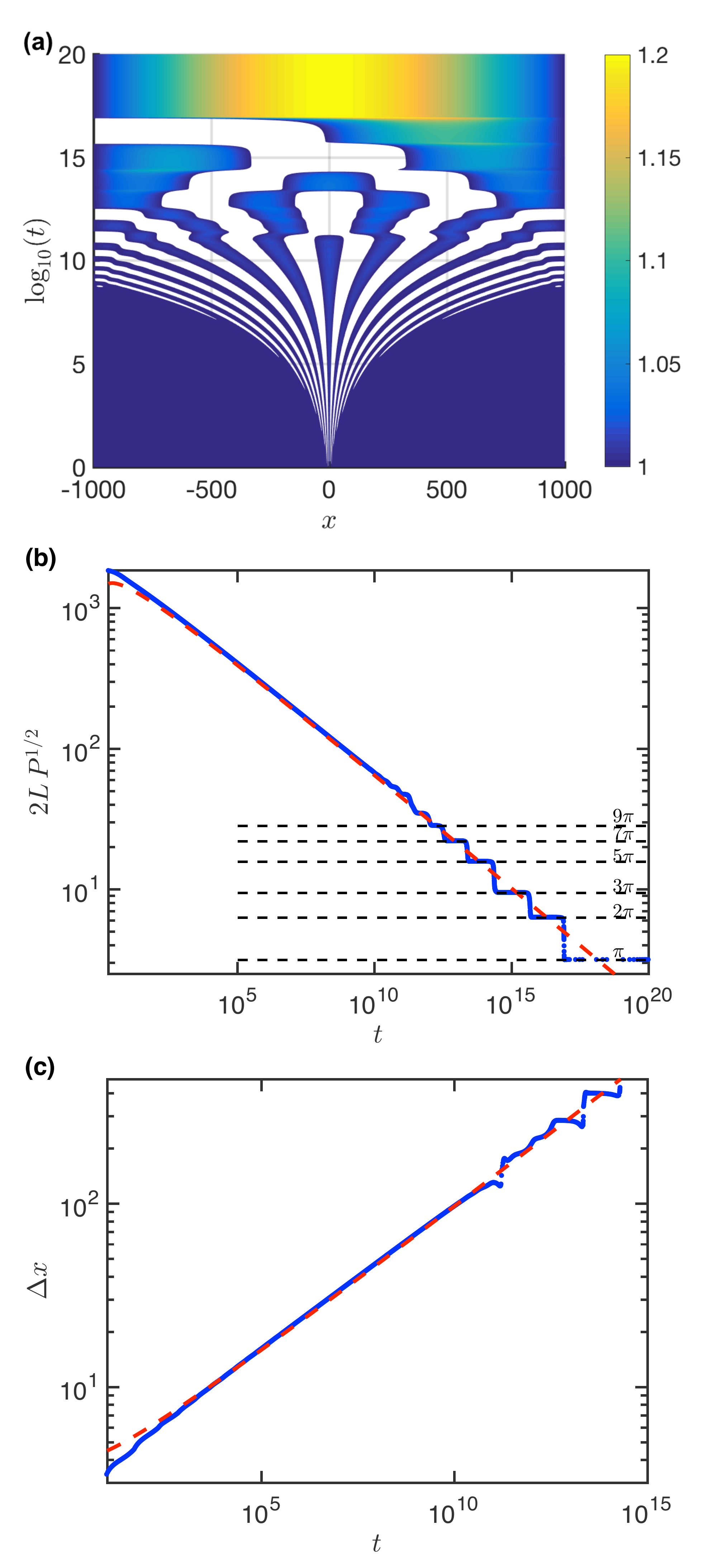}
  \caption{Numerical solution of the fully nonlinear system \eqref{eq:systembegin}--\eqref{eq:systemend} subject to a localized initial condition $u_0(x)=\varepsilon e^{-x^2}$. (a) The profile quickly develops wrinkles that eventually fill the entire domain before individually being squeezed out to give the final Euler-buckled mode. Colours represent the value of the film thickness, $h$, at each point of $(x,t)$-space, according to the colour bar on the right of the plot. Note that to highlight wrinkles, regions with $h(x,t)<1$ are whited-out. (b) The compressive force evolves according to the prediction of the linearized analysis \eqref{eq:Pasym} but ultimately breaks down when the presence of the boundaries becomes important.  For
    very late times the compressive force steps through periods where it is approximately constant, taking values close
    to the eigenvalues of the corresponding Euler-buckling problem, $2L P^{1/2}=(2n+1)\pi $ for $n=0,1,2,...$
    (indicated by horizontal dashed lines). (c)  The numerically determined distance between the first and the second zeros (points) is consistent with the asymptotic prediction \eqref{eqn:DeltaX} (dashed line). For very late times ( $t\gtrsim10^{15}$) only a few, larger bumps remain; at this stage the effect of the edges become important and the asymptotic result breaks down. In all simulations reported here, $L=1000$, $N=1024$, $\Delta = 4.9 \times 10^{-5}$ and $\varepsilon=10^{-2}$.}
\label{fig:4}
\end{figure}

The asymptotic analysis relies on the initial condition being localized (for the error in the application of Laplace's method to be small). In fig.~\ref{fig:4} we therefore show numerical results for a localized initial condition,  $u_0(x)=\varepsilon e^{-x^2}$ with $\varepsilon=10^{-2}$.  One might expect such an initial condition to yield a single bump spreading outwards; instead, the space--time plot of fig.~\ref{fig:4}a shows that wrinkles form on either side of the bump, resulting in a spreading corrugated pattern  a wavelength that increases with time. When the wrinkles reach the boundary this continuous gradual coarsening is replaced by quick jumps between the quasi-static Euler-buckling modes as the number of wrinkles decreases towards the final Euler-buckled state. We  see in fig.~\ref{fig:4}b that the compressive force $P(t)$ follows the asymptotic prediction \eqref{eq:Pasym} well, until the system starts to feel its size: the compressive force $P$ then gets temporarily `stuck' at a value that corresponds to one of the appropriate Euler-buckling modes. The system then transitions quickly to the next Euler buckling mode, as signified by a sudden decrease in $P$. The values of $P$ selected by our simulations at these very late times appears to depend on the symmetry of the initial condition: the symmetric initial condition used leads to values of $P$ close to those of the symmetric Euler buckling modes (i.e.~$L P^{1/2}=(n+1/2)\pi$, with $n$ an integer) until a single bump remains.

\subsubsection{A random initial condition}

\begin{figure}
  \centering
 \includegraphics[width=0.9\linewidth]{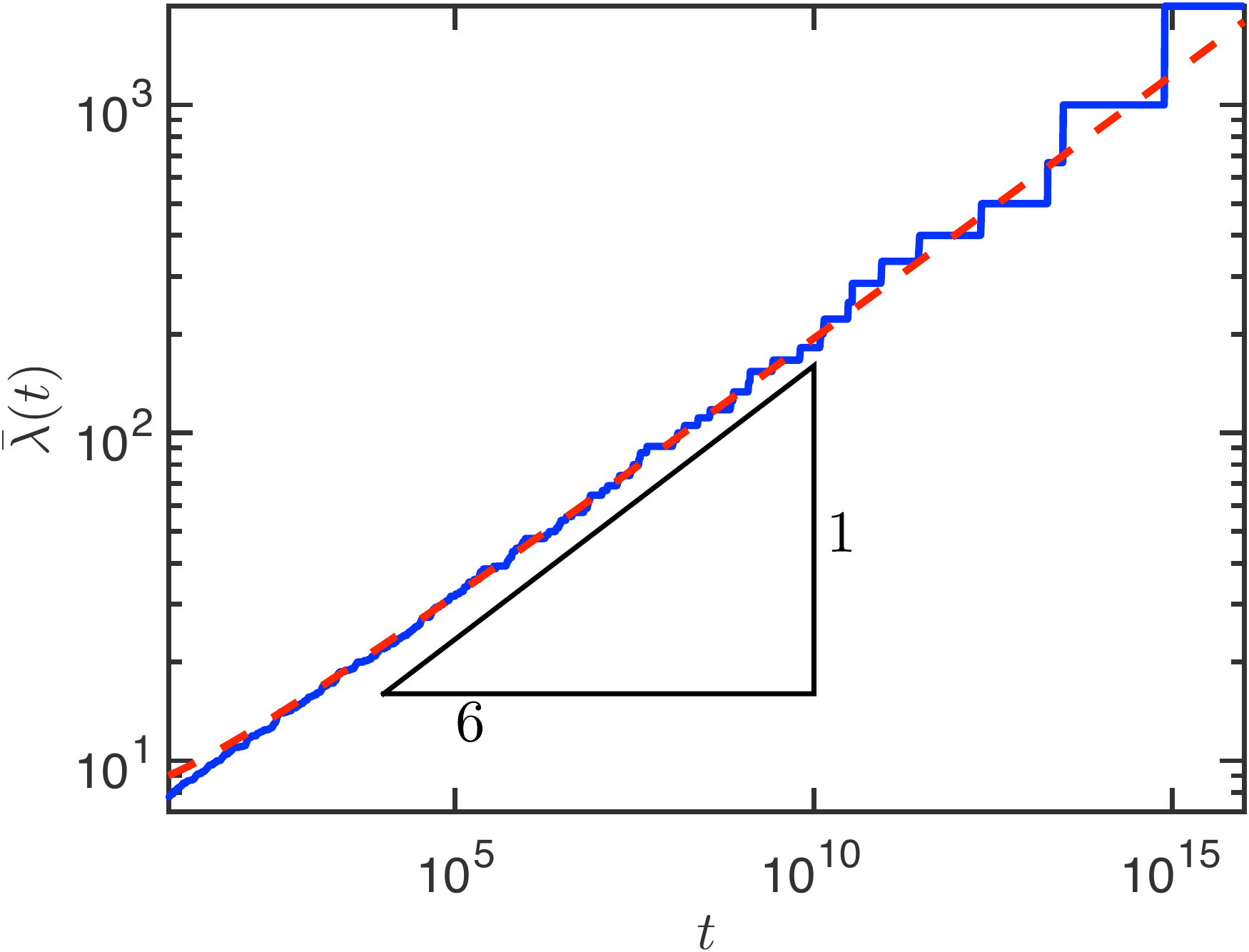}
  \caption{Wrinkle coarsening from a random initial condition appears to be governed by the evolution of localized bumps. Here the fully nonlinear problem is solved numerically with a randomized initial condition. The mean wrinkle wavelength ($\lbar=2L/n(t)$) evolves according to the long-time asymptotic prediction of the linearized problem with a localized initial condition, \eqref{eqn:LambdaBarOfT}, which is shown as the dash--dotted curve. The numerical results presented here are obtained with $L=1000$, $N=1024$, $\Delta = 17.7\times 10^{-3}$ and a random initial condition drawn from the uniform distribution on $[-\varepsilon,\varepsilon]$ with $\varepsilon=10^{-2}$. Note that at very late times the system gets temporarily stuck close to  Euler-buckled modes with sudden jumps between them; this accounts for the sudden large jumps in $\bar{\lambda}$ that are observed at late times. }
  \label{fig:5}
\end{figure}

It is also informative to examine data for the average wavelength in the fully nonlinear system with a random initial condition. A direct comparison of this numerical data and the corresponding asymptotic prediction \eqref{eqn:LambdaBarOfT} is shown in fig.~\ref{fig:5}. Again, this shows good agreement between the fully nonlinear problem and the linear analysis, at least until the wrinkles fill the domain. This is somewhat surprising since the initial condition used here is random and hence may not have a Fourier Transform that is dominated by the value of $\hat{u}_0(0)$, as is required for the late-time analysis to hold. This good agreement suggests that the existence of local inhomogeneities in the initial condition may be important in determining the evolution of the coarsening of wrinkles, as has been observed in elastocapillary aggregation  \cite{Singh2014}. As expected, the higher-order correction offered by including $\tauc$ improves the agreement between the asymptotic and numerical results (fig.~\ref{fig:5}).

\subsection{Comparison with previous experiments}

Having thoroughly investigated the evolution of the mean wrinkle wavelength using  numerical simulations and a linearized
analysis, we now return to reconsider the earlier experiments \cite{Vandeparre2010} that motivated this study. In their
experiments, Vandeparre \textit{et al.}~\cite{Vandeparre2010} studied the evolution of a thin titanium sheet (thickness  $\hTi =
15\mathrm{~nm}$, and Young's modulus $\ETi=10^{11}\mathrm{~Pa}$) above a thin substrate of polystyrene (thickness $h_{0}=250
\mathrm{~nm}$).  The  titanium--polystyrene composite is prepared as a solid and then heated above the  glass transition
temperature of the polystyrene. At  early times, the system wrinkles  with a wavelength determined purely by the elastic
properties of the system \cite{Stafford2004}. At later times  the polystyrene relaxes viscously and the wavelength starts to
grow as a function of time (all while the temperature remains fixed). Using different temperatures above the glass transition
temperature allowed the effective viscosity of the polystyrene to be varied, and hence the time scale of the evolution to be
varied too. However, the results as presented by Vandeparre \emph{et al.} \cite{Vandeparre2010}  are shifted to obtain a master curve at a temperature
close to $120\mathrm{^\circ C}$ using the WLF time--temperature superposition. 

\begin{figure}
  \centering
\includegraphics[width=\linewidth]{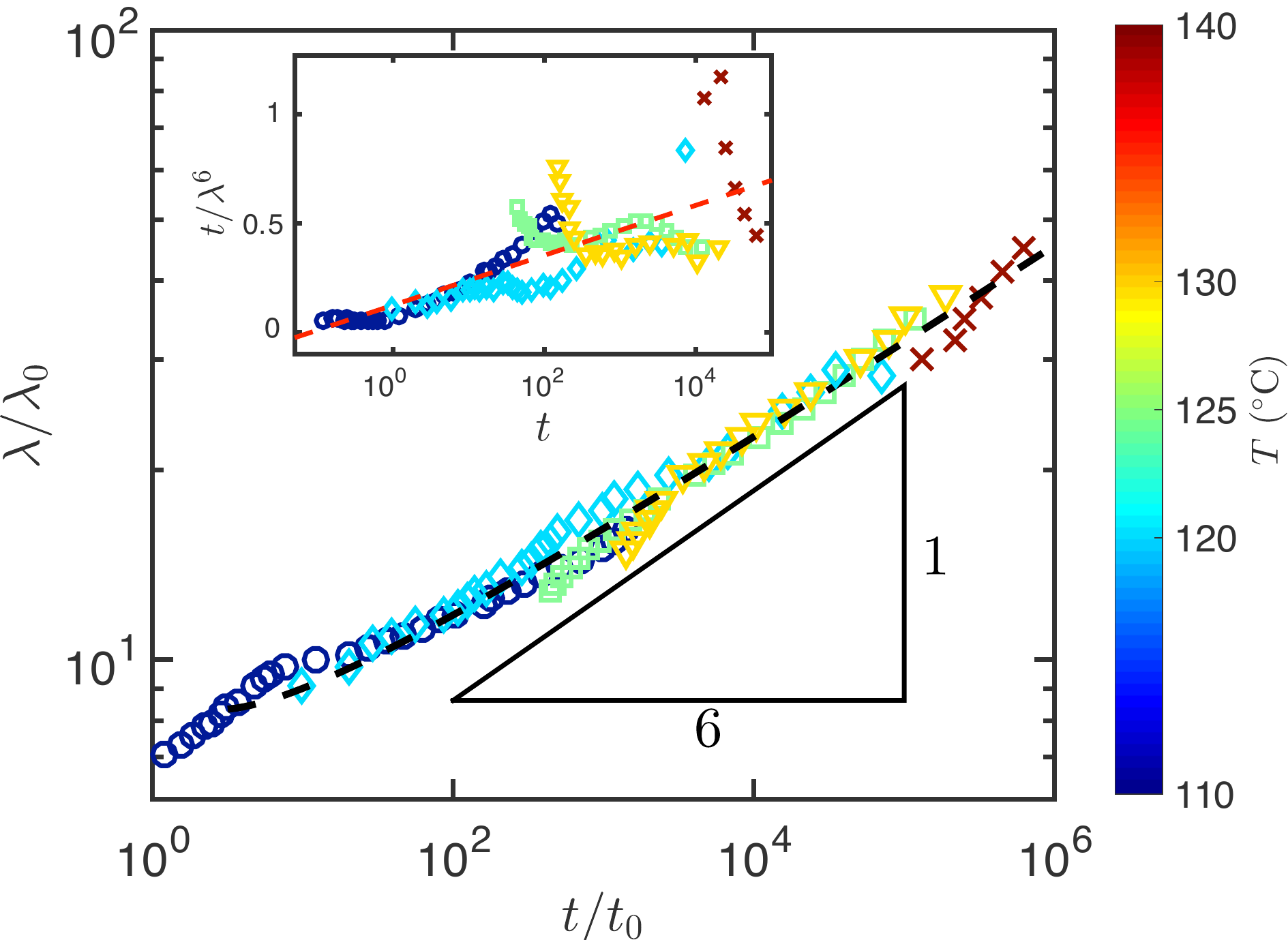}
  \caption{Comparison of previous experimental data \cite{Vandeparre2010} (points, courtesy of Pascal Damman) with the asymptotic prediction \eqref{eqn:LambdaBarOfT} (dashed curve).
Inset: Compensated plot of $t/\lambda^{6}$ which increases as a function of time instead of being a constant, as predicted by the naive scaling. Our model suggests that $t/\lambda^{6} \propto \log t/t_{c}$. We extract $t_{c}$ as the value of the time $t$ for which the best fit of $t/\lambda^{6}$ vanishes (red dashed line in the inset), giving $t_{c}= 0.0837 \mathrm{~s}$. In the main figure we non-dimensionalize  time by $t_{0}= t_{c}$ and lengths by $\lambda_{0} =\left[ B h_{0}^{3} t_{0}/(12 \mu) \right]^{1/6}\approx 0.16 \mathrm{~\mu m}$ where $\mu$ is used as a fitting parameter (within the limits given previously \cite{Plazek1971}). Experiments at different temperatures are  signified by different colours (as in the colour bar), and also using symbols: $T=110\mathrm{^\circ C}$ (circles), $T=120\mathrm{^\circ C}$ (diamonds), $T=125\mathrm{^\circ C}$ (squares), $T=130\mathrm{^\circ C}$ (triangles) and $T=140\mathrm{^\circ C}$ (crosses). These results have been shifted to a single effective temperature of $\approx120^\circ \mathrm{C}$ using the  WLF time-temperature superposition. 
}  
  \label{fig:6}
\end{figure}

Motivated by our asymptotic analysis, we re-examine the experimental data presented by Vandeparre \emph{et al.} \cite{Vandeparre2010}. In the inset of fig.~\ref{fig:6} we show a compensated plot of $t/\lambda^6$ versus $t$ (on a semi-logarithmic scale). This shows, firstly, that there is a systematic difference between experiments and the $\lambda\sim t^{1/6}$ behaviour expected from the naive scaling analysis (though we note that calculating $\lambda^6$ is likely to accentuate errors). Secondly, plotted in this way, the experimental data suggest the presence of a $\log t$ behaviour, as predicted by our theory \eqref{eqn:LambdaBarOfT}. The noise inherent in the data make it difficult to infer an effective value of $t_c$, and there are not sufficient experimental details to compute $t_c$. Nevertheless, the time scale $t_{c}$ is the time at which $t/\lambda^{6}$ vanishes on a semi-logarithmic plot (see inset of  fig \ref{fig:6}). Here, we  estimate this value to be $t_{c}= 0.0837 \mathrm{~s}$ from a best fit of all experimental data. We therefore non-dimensionalize times with a time scale $t_0=t_c$ and note that this corresponds to a length scale $\lambda_{0} =\left[ B h_{0}^{3} t_{0}/(12 \mu) \right]^{1/6}$. The length scale $\lambda_0$ can only be estimated once the viscosity $\mu$ is known. However, for the range of temperatures studied experimentally, estimates of $\mu$ vary in the range $10^{5}-10^{6} \mathrm{~Pa\, s}$ \cite{Plazek1971}. We therefore use the viscosity $\mu$ as a (the only) fitting parameter, finding that $\mu=2\times10^5\mathrm{~Pa\,s}$ (consistent with previous published data \cite{Plazek1971}); using standard values for titanium this  gives that $\lambda_0\approx 0.16\mathrm{~\mu m}$.

In figure \ref{fig:6} we compare the evolution of the wrinkle wavelength measured experimentally by Vandeparre \emph{et
  al.}~\cite{Vandeparre2010} with the long-time asymptotic behaviour predicted here, \eqref{eqn:LambdaBarOfT}. We see that the
asymptotic prediction \eqref{eqn:LambdaBarOfT} gives extremely good agreement with experiments, particularly for $t/t_0\gg1$. We
emphasize that in plotting fig.~\ref{fig:6} we have  fitted only one parameter  (the liquid viscosity $\mu$), with the same
value used to plot all data sets. Of particular interest is that for the time scales of experimental interest,  the difference
between the scaling prediction $t^{1/6}$ of Vandeparre \textit{et. al.}\cite{Vandeparre2010} and our result $(t/\log t)^{1/6}$ is relatively large: the line of best fit suggested previously \cite{Vandeparre2010} corresponds to an exponent $\approx0.133$. As a result we conclude that the refined theory presented here is likely to be of considerable use for understanding the moderate time scales that are accessible experimentally, for which the variation due to the $\log t$ is large enough to be observable.

\section{Conclusion}
\label{sec:conclusion}

In this paper, we have studied in detail the coarsening of wrinkles in a thin elastic beam on a thin viscous layer, subject to a constant end-shortening. Using a combination of numerical and asymptotic techniques we have shown that the nonlinear constraint of a fixed end--end compression modifies the behaviour of the system in important ways. Rather than a horizontal length scale $\propto t^{1/6}$, as has been frequently observed in analogous unconstrained problems \cite{Tulchinsky2016a,Huang2006}, we observe that the appropriate length scale $\propto (t/\log t)^{1/6}$. This logarithmic correction means that the evolution is no longer self-similar, though we are able to express the evolution of the beam profile in a form that is close to self-similar in the long-time limit.  This logarithmic correction results from the asymptotic evaluation of an integral (see  \eqref{eqn:FTconstraintAsy}) and so, to our knowledge, cannot be rationalized by means of a simple scaling argument. Nevertheless, we suggest that previous experimental measurements of an effective wavelength in fact show a discrepancy with the expected $t^{1/6}$ behaviour that is consistent with our prediction of $(t/\log t)^{1/6}$ behaviour.  As a result, even though logarithmic corrections are often ignored, they may be especially noticeable on the intermediate time scales of evolution that are accessible experimentally.

\acknowledgments

The research leading to these results has received funding from the European Research Council under the European Union's Horizon 2020 Programme / ERC Grant Agreement no.~637334 (DV) and from the Royal Society via a University Research Fellowship (IMG). We are grateful to Benny Davidovitch for bringing reference \cite{Vandeparre2010} to our attention,  to Pascal Damman for sharing the experimental data \cite{Vandeparre2010}, and to an anonymous reviewer for suggesting the analysis in Appendix \ref{app:NoFluid}. 

\begin{appendix}
\section{Numerical simulations}
\label{app:numerics}

To solve the system of equations  \eqref{eq:systembegin}--\eqref{eq:systemend} numerically, we discretize the spatial domain into $x_{i}$, $1\le i\le N+1$, in a way that conserves the flux on the interval $[(x_1+x_2)/2, (x_N,x_{N+1})/2]$. The end points are $x_{1}$ and $x_{N+1}$ and there are $N-1$ unknowns, $u_{i}=u(x_i)$, for $i=2,\ \dots,\ N$. The evolution of the $u_i$ is given by

\begin{align}
\label{eq:discretizedPDE}
&\frac{\upd u_i}{\upd t} = a_i p_{i+1} -(a_i + b_i) p_i +b_i p_{i-1}, \\
&p_i = \kappa_{i+1} -2 \kappa_i + \kappa_{i-1} + P \kappa_{i},\\
& \kappa_{i} = u_{i+1} -2 u_{i} + u_{i-1} , \\
& a_i = \frac{1}{\Delta x^{6}} [1+ (u_{i+1} +u_{i})/2]^{3} ,\\
& b_i = \frac{1}{\Delta x^{6}}   [1+ (u_{i} +u_{i-1})/2]^{3}, 
\end{align} 
subject to the imposed constraint \eqref{eq:constraintND}, which becomes
\begin{equation}
  \Delta = \frac{1}{2 \Delta x} \sum_{i=1}^{N} (u_{i+1} -u_{i})^{2}.
  \label{eq:discretizedConstraint}
\end{equation}
In order to obtain \eqref{eq:discretizedConstraint}, the integrand of the constraint, \eqref{eq:constraintND}, is computed using finite differences, centred at the half points $(x_i+x_{i+1})/2$.

The system \eqref{eq:discretizedPDE}--\eqref{eq:discretizedConstraint} is a differential algebraic equation (DAE) of index 2. We decrease its index to 1 by differentiating the constraint. The resulting system is a differential algebraic equation that may be integrated in time using the MATLAB routine ode15s (ode15s is able to handle a DAE of index 1 automatically). The solution of this DAE conserves the value of $\Delta$ of the initial condition to within $1\%$ in the simulations reported here.

\section{The breakdown of similarity solutions\label{app:SimSoln}}

From the simple scaling law presented previously \cite{Vandeparre2010} it is natural to assume that the partial
differential equation \eqref{eq:linearbegin} subject to the constraint \eqref{eq:linearend} should have a similarity
solution in which $x\sim t^{1/6}$, $u\sim t^{1/12}$, and $P\sim t^{-1/3}$. This expectation is further reinforced by the
calculation of such similarity solutions for the unconstrained problem \cite{Tulchinsky2016a}. However, we see from \eqref{eqn:uxtasym} that the true long-time solution for $u(x,t)$ cannot be expressed in similarity
form. For another perspective on why this approach does not work for the constrained problem considered here, we follow
the approach of Budd \emph{et al.} in a related problem \cite{Budd2000} and make the similarity ansatz \beq \eta =
x/t^{1/6},\quad u(x,t ) = t^{1/12} f(\eta),\quad P = t^{-1/3} Q.  \eeq Substituting this into \eqref{eq:linearend}, we
find that the governing partial differential equation becomes \beq
 \label{eq:similarity}
\frac{\upd^6 f}{\upd \eta^6} + Q \frac{\upd^4f}{\upd \eta^4}=\frac{1}{12} f -\frac{1}{6} \eta \frac{\upd f}{\upd \eta},
\eeq which is to be solved with the constraint
\beq
\frac{1}{2}\int_{-\infty}^\infty(f')^2~\upd\eta=\Delta.
\label{eqn:app:const}
\eeq

To see why the constraint \eqref{eqn:app:const} is incompatible with the similarity problem \eqref{eq:similarity}, we examine the behaviour of $f(\eta)$ as $\eta\to\infty$. This is most readily done by examining the $\omega\to0$ limit of the Fourier Transform $\hat{f}(\omega)$; we find that $\hat{f}(\omega)\propto \omega^{-3/2}$ as $\omega\to0$. However, in Fourier space, the constraint \eqref{eqn:app:const} reads
\beq
\tfrac{1}{2}\int_{-\infty}^\infty\omega^2\hat{f}^2~\upd\omega=\Delta.
\label{eqn:FT:const}
\eeq Since $\hat{f}(\omega)\propto \omega^{-3/2}$ as $\omega\to0$, the integral on the LHS of \eqref{eqn:FT:const} does not converge; we conclude that the solution of the similarity equation \eqref{eq:similarity} is not sufficiently well behaved as $\eta\to\pm\infty$ for the integral on the LHS of \eqref{eqn:app:const} to converge.

\section{Initial conditions with no excess of fluid, $\hat{u}_{0}(0) = 0$ \label{app:NoFluid}}

The asymptotic analysis presented in \S\ref{sub-sec-late-time} relied on the assumption that $\hat{u}_0(0)\neq0$, i.e.~that the initial condition contains an excess (or deficit) of fluid since $ \int_{-\infty}^\infty  u_{0}(x)~\upd x=\sqrt{2\pi}\hat{u}_{0}(0)\neq0$. 

\begin{figure}
  \centering
 \includegraphics[width=0.9\linewidth]{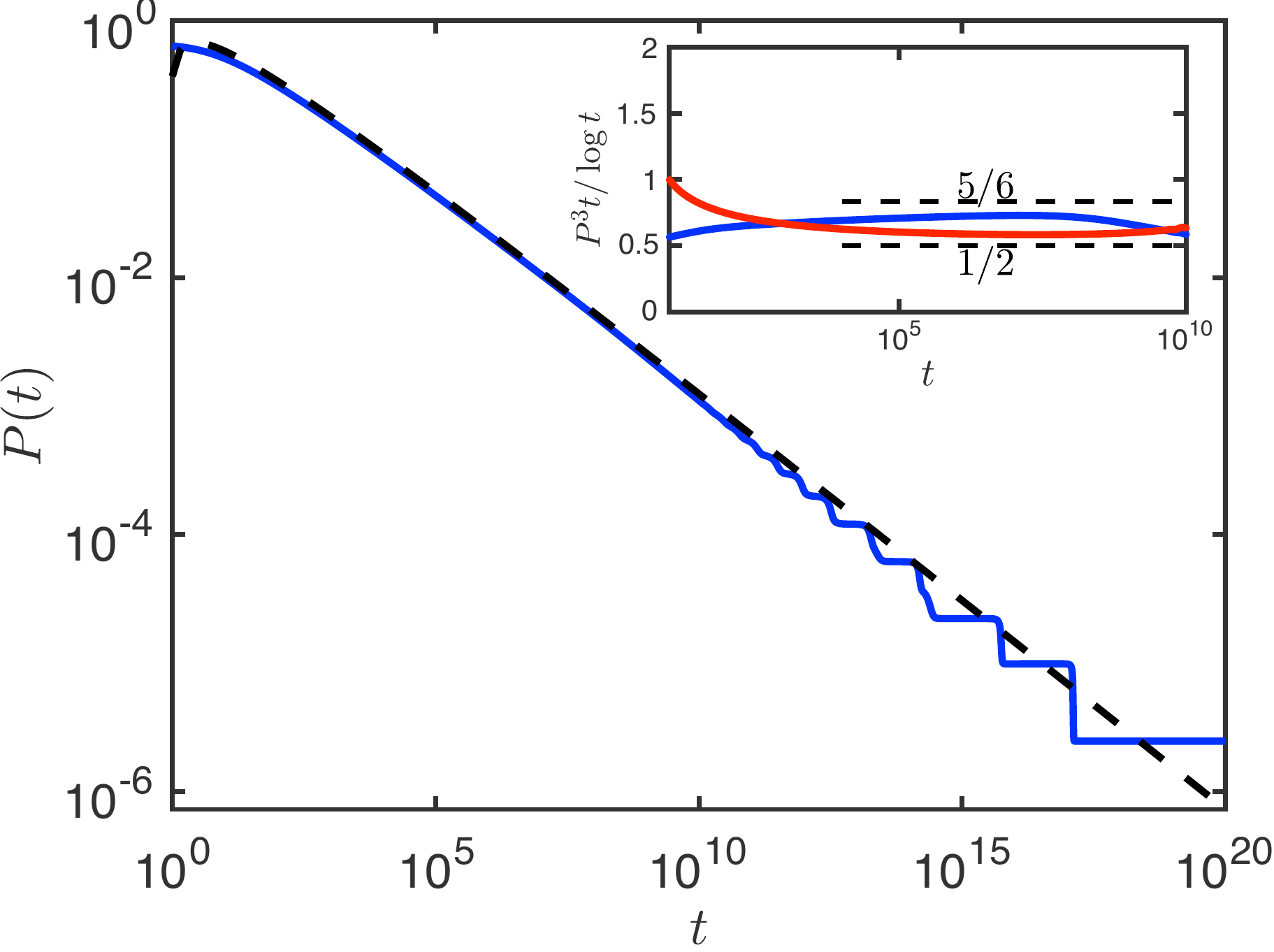}
  \caption{Validation of the asymptotic predictions for the case of no excess or deficit fluid, $\hat{u}_0(0)=0$. Numerical results are obtained  by solving equations
    \eqref{eq:systembegin}--\eqref{eq:systemend}. The blue curve represents the numerically determined
    compressive force, $P(t)$, and the black line is the asymptotic prediction \eqref{eq:P-anti-sym}.  The inset
    illustrates the compensated plot $P^{3}t/\log t$ which approaches the expected value of $5/6$ for the anti-symmetric case (blue curve) and $1/2$ for the symmetric case (red curve, data from fig.~\ref{fig:4}). 
The deviation at late times is due to the finite size of the domain.  Here the anti-symmetric initial condition is $u(x,0)=
\varepsilon x\exp(-x^2/4)$. The simulation is performed with $L=1000$, $N=1024$, $\varepsilon = 10^{-2}$, $\Delta=5\times10^{-5}$.}
\label{fig:7}
\end{figure}

Here, we consider the case in which the initial condition which has no deficit or excess of liquid so that $\hat{u}_0(0)=0$.  This may happen with a perfectly anti-symmetric initial condition.  The analysis follows through in much the same was as in \S \ref{sub-sec-late-time} but using $\hat{u}_0(k)\approx k\hat{u}_0'(0)$ for $k\ll1$ where
\begin{equation} 
 \hat{u}_0'(0) = -\frac{\mathrm{i}}{\sqrt{2\pi} }\int_{-\infty}^\infty u_0(x) x\,\mathrm{d}x 
\end{equation}
(so that it is the dipole moment of the  initial condition that dominates). We find that $\sigma \sim \frac{3^3}{2^3} \log[(t/t_{c})^{5/6}]$ at late times, where now $t_{c} = \left( \sqrt{3\pi} |\hat{u}' _{0}(0)|^{2}/9 \Delta \right)^{6/5}$ and so the late-time behaviour of the compressive force is
\begin{equation}
\label{eq:P-anti-sym}
P(t) \sim  \left(\tfrac{5}{6}\right)^{1/3} \left( \frac{\log t}{t} \right)^{1/3}.
\end{equation} 
Note that the prefactor in this new late-time behaviour of the compressive force eqn \eqref{eq:P-anti-sym} is $\left(5/6\right)^{1/3}\approx 0.941$, compared to $\left(1/2\right)^{1/3}\approx 0.794$ for an initial condition with an excess or deficit, $\hat{u}_0(0)\neq0$. To appreciate this difference between the two cases, the inset of fig.~\ref{fig:7} shows this difference via a compensated plot of the evolution of $P(t)$ in the two cases. 

At late times, the beam profile is given by 
\begin{equation}
  \label{eq:u-anti-sym}
u(x,t) \sim A_{1} \exp\left(-\frac{3}{32} \frac{\xi^{2}}{\sigma}\right) \left[ \sin(\sqrt{\tfrac{2}{3}}\xi) \right] ,
\end{equation}
where $A_{1} = \pm 2^{1/2} 3^{-1/4} \pi^{-1/4} \Delta^{1/2} t^{1/12} (\log t^{5/6})^{-1/3} $.

\end{appendix}

\end{document}